\documentclass[12pt, letterpaper]{article}

\usepackage[nottoc,notlot,notlof]{tocbibind}
\usepackage[title,titletoc]{appendix}

\usepackage{amsmath}
\usepackage{amsfonts}
\usepackage{amssymb}
\usepackage{graphicx}

\usepackage{color}

\usepackage{amsmath, amssymb, graphics, epsfig, graphicx}
\usepackage{epsf}
\usepackage{epstopdf}
\usepackage {amssymb}
\newcommand{\nc}{\newcommand}
\nc{\ba}{\begin{eqnarray}}
\nc{\ea}{\end{eqnarray}}
\newcommand\be{\begin{equation}}
\newcommand\ee{\end{equation}}

\newcommand{\calR}{{\cal{R}}}

\newcommand{\bfx}{{\bf{x}}}

\begin{document}

\vspace{5mm}
\vspace{0.5cm}
\begin{center}

\def\thefootnote{\fnsymbol{footnote}}

{\bf\large Two-field disformal transformation and mimetic cosmology}
\\[0.5cm]

{ Hassan Firouzjahi$^{1}\footnote{firouz@ipm.ir }$,  Mohammad Ali Gorji$^{1}\footnote{gorji@ipm.ir}$,    
Seyed Ali Hosseini Mansoori$^{ 2}\footnote{ shosseini@shahroodut.ac.ir, shossein@ipm.ir  }$, \\
Asieh Karami$^{1}\footnote{karami@ipm.ir}$, Tahereh Rostami$^{1}\footnote{t.rostami@ipm.ir }$,
}
\\[0.5cm]

{\small \textit{$^1$School of Astronomy, Institute for Research in Fundamental Sciences (IPM) \\ P.~O.~Box 19395-5531, Tehran, Iran
}}\\

{\small \textit{$^2$
Faculty of Physics, Shahrood University of Technology,\\ P.O. Box 3619995161 Shahrood, Iran }}

\end{center}

\vspace{.8cm}

\hrule \vspace{0.3cm}


\begin{abstract}

We extend the disformal transformation to models with two scalar fields and look at its 
singular limit. Solving the eigentensor equation for the Jacobian of the transformation 
of the metrics  we find the two-field extension of the mimetic scenario in the singular 
conformal  limit. At the background level the setup mimics the roles of dark matter 
cosmology. We decompose the perturbations into the adiabatic and entropy modes in 
which the adiabatic perturbation is tangential to the classical trajectory while the 
entropy mode is perpendicular to it. We show that the adiabatic mode is frozen while  
the entropy mode propagates with the sound speed equal to unity with no instabilities.

\end{abstract}
\vspace{0.5cm} \hrule
\def\thefootnote{\arabic{footnote}}
\setcounter{footnote}{0}
\newpage
\section{Introduction}

The scalar-tensor theories such as the Brans-Dicke \cite{Brans-Dicke}, Dirac-Born-Infeld 
\cite{Born:1934gh} and Horndeski \cite{Horndeski:1974wa} models are introduced to include 
a scalar field which makes the longitudinal mode of gravity dynamical. The scalar field can 
play the roles of dark matter or dark energy in late time cosmology and even inflaton in 
early universe cosmology. In the original formulation of Brans-Dicke model, the scalar field 
 couples non-minimally to the curvature and it does not couple to the matter sector. It turns 
out that the model can be rewritten as a standard Einstein-Hilbert action, {\it i.e.} without 
any coupling between the scalar field and the curvature, but now the scalar field couples 
non-minimally with the matter sector. The first frame is known as the Jordan frame and the 
latter is called the Einstein frame. These two frames are related to each other through a 
conformal transformation $g_{\mu\nu}\rightarrow f(\phi) g_{\mu\nu}$ and they are equivalent 
at the classical level \cite{Flanagan:2004bz, Deruelle:2010ht, Chiba:2013mha}. 

In the Dirac-Born-Infeld model, the derivative of scalar field couples non-minimally to the 
curvature which can be eliminated through a disformal transformation \cite{Zumalacarregui:2012us}
\be\label{disformal-0}
g_{\mu\nu}\rightarrow A(\phi) g_{\mu\nu} + B(\phi) \phi_{,\mu}\phi_{,\nu}\,,
\ee
in which we have used the notation $\phi_{,\mu}\equiv\partial_{\mu}\phi$. The Horndeski theories 
are the most general scalar tensor models which include the higher derivatives of scalar field in 
the action while the equations of motion remain second order and therefore they are free
of the so-called Ostrogradsky ghost \cite{Horndeski:1974wa}. Applying disformal 
transformation (\ref{disformal-0}) to the Horndeski models, it is shown that the transformed
models still belong to Horndeski models through appropriate redefinition of the coefficients
\cite{Bettoni:2013diz}. 

Considering the more general disformal transformation
\be\label{disformal-1}
g_{\mu\nu}\rightarrow A(\phi) g_{\mu\nu} + B(\phi,X) \phi_{,\mu}\phi_{,\nu}\,,
\ee
in which $X\equiv g^{\mu\nu}\phi_{,\mu}\phi_{,\nu}$, the Horndeski model is converted into the beyond
Horndeski model in which the equation of motion is no longer second order while the setup is 
still free of the Ostrogradsky ghost \cite{disformal-beyondHorndeski}. In this respect, one tempts 
to consider the most general disformal transformation of the form 
\be\label{disformal}
g_{\mu\nu}\rightarrow A(\phi,X) g_{\mu\nu} + B(\phi,X) \phi_{,\mu}\phi_{,\nu}\,,
\ee
which was first suggested by Bekenstein \cite{Bekenstein:1992pj}. Applying the above general 
disformal transformation to the Horndeski model, one again finds higher derivative models which
are free of  Ostrogradsky ghost \cite{Zumalacarregui:2013pma}. In this respect, the disformal 
transformations reveal that the equation of motion is not a fundamental criterion to avoid 
the Ostrogradsky ghost. Recently, in an interesting paper \cite{Langlois:2015cwa}, it is shown 
that the degeneracy of all these models make them free of Ostrogradsky ghost even in the 
presence of higher derivative terms and higher derivative equations of motion. Such theories, 
known as DHOST (Degenerate Higher Order Scalar Tensor), are the most general theories which 
include  the second derivatives of the scalar field in the action 
while are free of the Ostrogradsky ghost. They are formally invariant under the general disformal 
transformation (\ref{disformal}) \cite{disformal-DHOST}. It is important to note that all of the 
above discussions are valid only in the absence of matter fields and these models are not 
invariant under the disformal transformation in the presence of matter sector.

Of more interests are singular disformal transformations where the number of degrees of freedom 
is no longer preserved between the two frames. One can find physically
different models by performing a singular disformal transformation to a well-known scalar-tensor
model. The simplest  example is the pure tensorial Einstein-Hilbert action in which under a singular transformation,  the longitudinal mode of gravity becomes dynamical in the new frame. This scenario is known as the mimetic dark matter in which the scalar field plays the roles of dark matter  \cite{Mimetic-2013}. The cosmological and theoretical aspects of the setup have been studied 
widely \cite{Mimetic}. The original mimetic scenario is free of pathologies but the scalar mode 
corresponding to the longitudinal mode of gravity is frozen at the level of linear perturbations 
\cite{Barvinsky:2013mea, Chaichian:2014qba}. In order to have a propagating scalar mode, it is 
suggested to add a higher derivative term to the action \cite{Mimetic-2014} such that the setup 
still describes dark matter at the level of background \cite{Mirzagholi:2014ifa}. This setup then 
turns out to be unstable even at the level of linear perturbations 
\cite{Ramazanov:2016xhp, Ijjas:2016pad, Firouzjahi:2017txv}. Finally, it is shown that the model 
can be stabilized by adding some coupling between the curvature and second derivatives of 
mimetic scalar field \cite{Hirano:2017zox, Zheng:2017qfs, Gorji:2017cai}. 

Here we look for the two-field extension of the original mimetic scenario. In order to do this, we note that the  
original single field mimetic scenario can be  realized from the singular limit of the disformal transformation 
Eq. (\ref{disformal}) in the conformal case $B=0$ \cite{Deruelle:2014zza, Arroja:2015wpa, Domenech:2015tca}. 
In this regard, we first find the two-field extension of the derivatively coupled disformal transformation 
Eq. (\ref{disformal}) and then look at its singular limit in the special case of conformal transformation.
Interestingly, we find that the two-field generalization of the mimetic scenario still describes 
a dark matter-like fluid in cosmological background and also it is free of disastrous pathologies 
and provides healthy entropy mode at the level of perturbations.

\section{Two-field Disformal Transformation}
\label{two-field-dis}

The natural generalization of the general disformal transformation Eq. (\ref{disformal}) to the case of
two scalar fields would have the following form
\be
\label{dis}
g_{\mu\nu}= A {\tilde g}_{\mu\nu}+B \, \phi_{,\mu}\phi_{,\nu} 
+ C\, \psi_{,\mu}\psi_{,\nu}
+ D\, (\phi_{,\mu}\psi_{,\nu}+\psi_{,\mu}\phi_{,\nu}) \, ,
\ee
where $A, B, C, D$ are given functions of $\phi, \psi, X, Y, Z$ where $X, Y, Z$ are defined as
\be
\left\{
\begin{array}{c}
	X \equiv {\tilde g}^{\mu\nu} \phi_{,\mu}\phi_{,\nu}\,, \\
	Y \equiv {\tilde g}^{\mu\nu} \psi_{,\mu}\psi_{,\nu}\,, \\
	Z \equiv {\tilde g}^{\mu\nu} \phi_{,\mu}\psi_{,\nu} \,.
\end{array}
\right.
\ee
One may consider $g_{\mu \nu}$ as the final ``physical" metric while ${\tilde g}_{\mu \nu}$ may 
be viewed as the initial ``auxiliary" metric.

Demanding that the determinant of $g_{\mu\nu}$ to be nonzero, we can seek for the inverse 
metric 
\be\label{inverse-metric}
{g}_{\mu\nu}{g}^{\mu\alpha} = \delta_{\nu}^\alpha \,.
\ee
We consider the following form for the inverse metric
\be\label{metric-inverse}
g^{\mu\nu}=\bar A {\tilde g}^{\mu\nu}+\bar B {\tilde g}^{\mu\alpha}{\tilde g}^{\nu\beta}
\phi_{,\alpha}\phi_{,\beta} + \bar C {\tilde g}^{\mu\alpha}{\tilde g}^{\nu\beta}
\psi_{,\alpha}\psi_{,\beta}
+\bar D {\tilde g}^{\mu\alpha}{\tilde g}^{\nu\beta}(\phi_{,\alpha}\psi_{,\beta}
+\psi_{,\alpha}\phi_{,\beta}) \,,
\ee
in which $\bar A, \bar B, \bar C, \bar D$ are unknown functions of $\phi, \psi, X, Y, Z$ and
our task is to find their explicit forms in terms of the known functions $A, B, C, D$ in Eq. (\ref{dis}).
Substituting Eq. (\ref{dis}) and the ansatz Eq. (\ref{metric-inverse}) into the relation 
Eq. (\ref{inverse-metric}), we find five equations for four undefined functions $\bar A, \bar B, 
\bar C, \bar D$. One of these equations turns out not to be independent and therefore we are left 
with four independent equations for four undefined functions, yielding the following solutions
\begin{eqnarray}\label{coefficient-solution}
&&\bar A=A^{-1} \,, \nonumber \\ 
&&\bar B=-\frac{A B+B C Y-D^2 Y}{A\,  T} \,, \nonumber \\
&&\bar C=-\frac{A C+B C X-D^2 X}{A \,  T} \,, \nonumber \\
&&\bar D=-\frac{A D-B C Z+D^2 Z}{A\,  T} \, ,
\end{eqnarray}
in which 
\ba
T\equiv A^2 + (B X + C Y + 2 D Z) A + ( X Y - Z^2) ( BC - D^2) \, .
\ea

A variant of two-field disformal transformation was performed in \cite{Yuan:2015tta} by applying two 
successive single field disformal transformation. However, the last ({\it cross}) term in Eq. (\ref{dis}) cannot 
be generated by two successive single field disformal transformations of the form Eq. (\ref{disformal}). 
Having said this, we show in the appendix \ref{app-diag} that this cross term can be eliminated by 
means of an appropriate linear map in cotangent bundle of the field space. Therefore, without loss of 
generality we may set $D=0$. However, note that even if one starts with no cross term in 
$g_{\mu \nu}$ by setting $D=0$, one still has the cross term $\bar D \neq 0$ in the inverse metric 
(\ref{metric-inverse}) as can been  from Eq. (\ref{coefficient-solution}).

To find if the transformation $g_{\mu\nu}\rightarrow\,{\tilde g}_{\mu\nu}$ is invertible or under what 
conditions we can obtain ${\tilde g}_{\mu\nu}={\tilde g}_{\mu\nu}(g_{\mu\nu})$, we should look at the Jacobian 
of the transformation $\frac{\partial g_{\mu\nu}}{\partial {\tilde g}_{\alpha\beta}}$. For the special case of 
two-field disformal transformation when the coefficients $A, B, C, D$ are only functions of $\phi$ and 
$\psi$, the Jacobian is simply given by $\frac{\partial g_{\mu\nu}}{\partial {\tilde g}_{\alpha\beta}}
=A\delta^\alpha_\mu \delta^\beta_\nu$. Therefore, as long as $A\neq0$, we can re-express the auxiliary 
metric ${\tilde g}_{\mu\nu}$ in terms of the physical metric $g_{\mu\nu}$ as ${\tilde g}_{\mu\nu} 
= \frac{1}{A} g_{\mu\nu} - \frac{B}{A} \, \phi_{,\mu}\phi_{,\nu} - \frac{C}{A}\, \psi_{,\mu}\psi_{,\nu} 
- \frac{D}{A}\, (\phi_{,\mu}\psi_{,\nu}+\psi_{,\mu}\phi_{,\nu})$. 

For the general case  where $A, B, C, D$ are functions of not only $\phi$ and $\psi$ but also $X, Y, Z$, 
then the Jacobian  $\frac{\partial g_{\mu\nu}}{\partial {\tilde g}_{\alpha\beta}}$ does not has a simple form. 
Equivalently, we  can look at the eigenvalue equation for the determinant of the Jacobian 
\cite{Zumalacarregui:2013pma}
\be
\label{eigen}
\left(\frac{\partial g_{\mu\nu}}{\partial {\tilde g}_{\alpha\beta}}-\lambda^{(n)}\,
\delta_\mu^\alpha\delta_\nu^\beta\right)\xi^{(n)}_{\alpha\beta}=0 \,,
\ee
where $\lambda^{(n)}$ are the eigenvalues and $\xi^{(n)}_{\mu\nu}$ are the associated 
eigentensors. Our task is now to find the eigenvalues which determine whether or not the disformal 
transformation (\ref{dis}) is invertible.


\subsection{Conformal case}

As we have already mentioned in the Introduction section, the original single field mimetic 
scenario can be realized from the singular conformal limit of (\ref{disformal}). Since we are 
interested in two fields extension of the mimetic scenario, in this subsection, we consider
the conformal case with $B=C=D=0$ in the general disformal transformation (\ref{dis}). The
general case is studied in the Appendix \ref{app-general-dis}. 

In the case of two-field conformal transformation, we have 
\be\label{conformal}
g_{\mu\nu} = A(\phi,\psi,X,Y,Z) \, {\tilde g}_{\mu\nu} \,.
\ee
Clearly, the inverse metric (\ref{metric-inverse}) takes the simple form of $g^{\mu\nu} = 
A^{-1} \, {\tilde g}^{\mu\nu}$ which can also be obtained from (\ref{coefficient-solution}) when
$B=C=D=0$.

The eigenvalues equation (\ref{eigen}) for the conformal transformation (\ref{conformal}) 
simplifies to
\be
\label{eigen-c}
(A-\lambda)\xi_{\mu\nu} - \Big(A_{,X} \langle \xi\rangle_{X}+A_{,Y} \langle \xi\rangle_{Y}
+A_{,Z} \langle \xi\rangle_{Z} \Big) g_{\mu\nu} = 0 \,,
\ee
where we have defined

\be\label{xis-def}
\langle\xi\rangle_{X} \equiv \xi_{\alpha\beta} \phi^{,\alpha} \phi^{,\beta} \,, 
\hspace{1cm}
\langle\xi\rangle_{Y} \equiv \xi_{\alpha\beta} \psi^{,\alpha} \psi^{,\beta} \,, 
\hspace{1cm}
\langle\xi\rangle_{Z} \equiv \xi_{\alpha\beta} \phi^{,\alpha} \psi^{,\beta} \,.
\ee

There are two types of solutions for the eigenvalue problem (\ref{eigen-c}) which we call 
``conformal type solution'' and ``kinetic type solution.'' 

The conformal type eigenvalue and the associated eigentensor are given by

\be\label{eigenvalue-Cphipsi}
\lambda^{C} = A \,, \hspace{1cm} \mbox{with}
\hspace{1cm} A_{,X} \langle \xi^C\rangle_{X}+A_{,Y} \langle \xi^C\rangle_{Y}+A_{,Z} \langle \xi^C\rangle_{Z} = 0  \,.
\ee
Note that the conformal type eigenvalue is degenerate with multiplicity of $9$ since the associated
eigentensors are restricted to the above (single) constraint.

From (\ref{eigen-c}), it is clear that the remaining kinetic type eigentensor will be proportional to the 
metric tensor and therefore we find
\be\label{eigenvalue-Kphipsi}
\lambda^{K} = A -\left(X A_{,X}+Y A_{,Y}+Z A_{,Z}\right)\,, \hspace{1cm} \mbox{with}
\hspace{1cm} \xi_{\mu\nu}^{K} = {\tilde g}_{\mu\nu} \,.
\ee
	
Having obtained the eigenvalues (\ref{eigenvalue-Cphipsi}) and (\ref{eigenvalue-Kphipsi}) we can easily 
find the singular limit of the two-field conformal transformation (\ref{conformal}) by demanding that 
the eigenvalues to vanish. Before doing this, let us elaborate more on the physical meaning of this condition.  
Indeed, in the case of an invertible transformation with non-vanishing eigenvalues (\ref{eigenvalue-Cphipsi}) 
and (\ref{eigenvalue-Kphipsi}), the scalar fields $\phi$ and $\psi$ do not play any significant roles though 
they appear explicitly in the action in the transformed frame. This is because  we can always perform the 
inverse transformation and remove all of the effects of $\phi$ and $\psi$. But, as we will see in the next 
Section, when the  transformation is singular we can not remove the effects of $\phi$ and $\psi$ through 
any field redefinition. More precisely, the number of physical degrees of freedom does not change under 
an invertible disformal/conformal transformation while it changes when the transformation is singular. In 
single field mimetic gravity  the number of degree of freedom increases  through singular conformal 
transformation. There are two tensor degrees of freedom  associated to the gravitons in the original 
(untransformed) frame while there are three degrees of freedom in the  transformed frame 
\cite{Deruelle:2014zza}. The longitudinal mode of gravity becomes dynamical in the transformed frame and 
an extra scalar mode appears. Correspondingly, in the two field case, we expect that the number of degrees 
of freedom does not change under regular conformal transformation (\ref{conformal}) (or disformal 
transformation (\ref{dis})). However,  it increases by two scalar degrees of freedom, corresponding to two 
scalar fields $\phi$ and $\psi$,  when the  conformal transformation (\ref{conformal}) is singular. 

In linear cosmological perturbations, the new scalar mode in the original single field mimetic scenario is not a propagating mode.  In other words, the curvature perturbation is frozen 
\cite{Mimetic-2014}. In correspondence, we also expect to find only one propagating scalar mode in our two-field extension of original mimetic scenario. We show that  the curvature perturbation is still frozen to linear order in cosmological perturbations  in our setup while there is one propagating entropy mode.


\section{Two-field Mimetic Dark Matter}

Now, let us consider the singular limit of conformal transformation (\ref{conformal}). For the 
conformal type solution (\ref{eigenvalue-Cphipsi}) this happens when $A=0$ which is not 
allowed.  For the kinetic type solution (\ref{eigenvalue-Kphipsi}), however, the singular limit 
exists with the following condition on $A$: 
\ba
\label{singular-A}
A= X A_{,X} + Y A_{,Y} + Z A_{,Z} \,  \quad \quad (\mathrm{singular\, \, limit}) \, .
\ea

We can obtain the two-field generalization of the mimetic scenario by looking at the singular 
limit of the two-field disformal transformation. The nontrivial solution for the conformal factor $A$ satisfying condition (\ref{singular-A}) is

\be\label{singular-solution}
A = - \alpha\, X - \beta\, Y - 2 \gamma\, Z \,.
\ee
where $\alpha$, $\beta$, and $\gamma$ are arbitrary functions of the scalar fields $\phi$ and
$\psi$ and the minus signs and the factor $2$ in front of the last term are considered for 
convenience.

In the appendix \ref{app-diag} we have shown that the cross term defined by $Z$ in Eq. 
(\ref{singular-solution}) can be removed through the linear transformation Eq. 
(\ref{transformation-cotangent}). Therefore, without loss of generality, we  can set 
$\gamma=0$ in the analysis below. 

Substituting Eq. (\ref{singular-solution}) in Eq. (\ref{conformal}), we find that the singular 
conformal transformation would have the following  form 
\be\label{metric-singular}
g_{\mu\nu} = - \big(\alpha\, {\tilde g}^{\sigma\rho}\phi_{\sigma}\phi_{\rho}
+ \beta\, {\tilde g}^{\sigma\rho}\psi_{\sigma}\psi_{\rho} \big) {\tilde g}_{\mu\nu} \,.
\ee

Note that we can not obtain ${\tilde g}_{\mu\nu}$ as a function of $g_{\mu\nu}$ which 
demonstrates the singular nature of the above transformation. It is also easy to check that 
the physical metric $g_{\mu\nu} $ is invariant under conformal transformation of the 
auxiliary metric ${\tilde g}_{\mu\nu}$. In addition, the inverse of the metric $g_{\mu\nu}$ 
from (\ref{metric-singular}) can be read off as
\be\label{metric-singular-inevrse}
g^{\mu\nu} = - \big(\alpha\, {\tilde g}^{\sigma\rho}\phi_{\sigma}\phi_{\rho}
+ \beta\, {\tilde g}^{\sigma\rho}\psi_{\sigma}\psi_{\rho} \big)^{-1} {\tilde g}^{\mu\nu} \,.
\ee

Contracting both sides of the above relation with $\phi_{\mu}\phi_{\nu}$ and 
$\psi_{\mu}\psi_{\nu}$ we obtain $g^{\mu\nu}\phi_{\mu}\phi_{\nu}= - X / (\alpha X + 
\beta Y)$ and $g^{\mu\nu}\psi_{\mu}\psi_{\nu}= - Y / (\alpha X + \beta Y)$ which 
implies

\be\label{mimetic-constraint}
\alpha(\phi,\psi) g^{\mu\nu}\phi_{\mu}\phi_{\nu} 
+ \beta(\phi,\psi) g^{\mu\nu}\psi_{\mu}\psi_{\nu} = -1 \,.
\ee

Therefore, the two-field conformal transformation (\ref{conformal}) is not invertible if the physical 
metric satisfies the above constraint. It is easy to see that all of the above results reduce to the case 
of single field mimetic scenario when $\beta=0$ ($\alpha=0$) and $\alpha=\mbox{const.}$ 
($\beta=\mbox{const.}$). The special case of $\alpha=\alpha(\phi)$ and $\beta=\beta(\psi)$ coincides 
with the model proposed in \cite{Vikman:2017gxs} if we neglect the electromagnetic field in that model.

The shift symmetry condition for the scalar fields is not necessary for the 
conformal transformation (\ref{conformal}) to be singular. However,  this condition is imposed in the original mimetic scenario  \cite{Mimetic-2013} in order to obtain a dark matter-like fluid. More precisely, as we will explicitly show  here, the existence of the Noether current associated with the shift symmetry provides a dark matter-like  energy density component at the cosmological background. Therefore, we assume the shift symmetry for both scalar fields $\phi$ and $\psi$. Moreover, instead of applying the singular transformation 
Eq. (\ref{metric-singular}) directly, it is convenient to include the constraint Eq. 
(\ref{mimetic-constraint}) into the action through a Lagrange multiplier \cite{Golovnev:2013jxa} (see 
also\cite{Lim:2010yk}) so that the action of our model takes the following form
\ba\label{action00}
S= \int d^4 x  \sqrt{-{ g}} \left[ \frac{1}{2} { R} + {\tilde \lambda} \left( 
\alpha { g}^{\mu\nu} \phi_{,\mu} \phi_{,\nu} +
\beta { g}^{\mu\nu} \psi_{,\mu} \psi_{,\nu} + 1 \right)\right] \,,
\ea
in which ${\tilde \lambda}$ is a Lagrange multiplier which enforces the mimetic constraint
(\ref{mimetic-constraint}). Note that the two functions $\alpha$ and $\beta$ are constant since we 
have imposed the shift symmetry for both scalar fields. Now, without loss of generality,  we can absorb constants $\alpha$ and $\beta$ into the fields through the field redefinitions $\phi
\rightarrow\phi/\sqrt{\alpha}$ and $\psi\rightarrow\psi/\sqrt{\beta}$. The ratio of these constants
$\beta/\alpha$, however, determines the relative contributions of each scalar field to the total kinetic 
term. In other words, it is plausible to expect that this ratio would be related to the entropy of the 
fluid that describes our model. Therefore, we absorb $\alpha$ in the auxiliary field through the field 
redefinition ${\tilde \lambda}\rightarrow\lambda/\alpha$ but keep $\beta$ by defining the constant 
parameter ${\cal S}=\sqrt{\beta/\alpha}$. If we set ${\cal S}$ to zero, all the effects of the extra field $\psi$ 
disappear and we find a single field mimetic scenario.  Therefore ${\cal S}$ would be related to the entropy in some sense. We will see that this is indeed 
the case and ${\cal S}$ is nothing but the constant value of the entropy field at the background. 

With these discussions,  the action (\ref{action00}) becomes
\ba\label{action0}
S= \int d^4 x  \sqrt{-{ g}} \left[ \frac{1}{2} { R} + \lambda \left( 
{ g}^{\mu\nu} \phi_{,\mu} \phi_{,\nu} +
{\cal S}^2 { g}^{\mu\nu} \psi_{,\mu} \psi_{,\nu} + 1 \right)\right] \,,
\ea
where $\lambda$ enforces the mimetic constraint 
\ba\label{mimetic-const.}
{ g}^{\mu\nu} \phi_{,\mu} \phi_{,\nu}
+ {\cal S}^2 { g}^{\mu\nu} \psi_{,\mu} \psi_{,\nu} = -1 \,.
\ea

Varying the action (\ref{action0}) with respect to the metric $g_{\mu\nu}$, one leads to the 
Einstein fields equations $G_{\mu\nu} = T_{\mu\nu} $ in which the effective energy momentum 
tensor is given by
\begin{equation}\label{Tmu-nu}
T^\mu_\nu=-2\lambda \left(\phi^{,\mu} \phi_{,\nu} 
+ {\cal S}^2 \psi^{,\mu} \psi_{,\nu} \right)\,.
\end{equation}
Note that the energy momentum tensor also contains a term of the form of
$\delta^\mu_\nu (g^{\alpha \beta } \phi_{,\alpha} \phi_{,\beta}
+ {\cal S}^2 g^{\alpha \beta } \psi_{,\alpha} \psi_{,\beta} + 1 )$ which vanishes after imposing  the 
constraint Eq. (\ref{mimetic-const.}).

In addition, varying the action with respect to $\phi$ and $\psi$ yields the modified 
Klein-Gordon equations
\begin{equation}\label{KG-eq}
\big( \sqrt{-g} \,\lambda\, \phi^{,\mu} \big)_{,\mu} = 0 , \hspace{1cm} 
\big( \sqrt{-g} \,\lambda\,  {\cal S}\psi^{,\mu} \big)_{,\mu} = 0 .
\end{equation}

The above equations show that the quantities $\sqrt{-g} \,\lambda\, \phi^{,\mu}$ and $\sqrt{-g} \,
\lambda\, {\cal S} \psi^{,\mu}$ are conserved. Indeed, these are nothing but the Noether currents associated
with the shift symmetries $\phi\rightarrow\phi+\mbox{const.}$ and $\psi\rightarrow\psi+\mbox{const.}$ 
respectively. Thus, we have
\begin{equation}\label{Noether-currents}
\phi^{,\mu} = \frac{\alpha^{\mu}}{\sqrt{-g} \,\lambda} , \hspace{1cm} 
{\cal S} \psi^{,\mu} = \frac{\beta^{\mu}}{\sqrt{-g} \,\lambda} ,
\end{equation}
where $\alpha^{\mu}$ and $\beta^{\mu}$ are constants of integrations. We assume that both 
$\phi^\mu$ and $\psi^\mu$ to be timelike and therefore the mimetic constraint (\ref{mimetic-const.})
is satisfied in either case of $\phi=0$ or $\psi=0$. In this respect, $\alpha^{\mu}$ and $\beta^{\mu}$
would be timelike vectors with constant components so that $\alpha_{\mu}\alpha^\mu<0$ and
$\beta_{\mu}\beta^{\mu}<0$. Substituting (\ref{Noether-currents}) into the mimetic constraint 
(\ref{mimetic-const.}) and then taking the square root, we we find 
\be\label{lambda-volume}
\lambda = \frac{{\sqrt{-(\alpha_\mu\alpha^\mu + \beta_{\mu}\beta^{\mu})}}}{\sqrt{-g}} \, .
\ee
From (\ref{Tmu-nu}), we see that the energy density is proportional to $\lambda$ and, as we shall see below from Eq. (\ref{lambda}), our setup provides an energy density component which behaves like dark 
matter at the cosmological background. Note that if we do not assume shift symmetry, then the derivatives of 
functions $\alpha$ and $\beta$ would appear in the right hand side of (\ref{KG-eq}) and therefore we could not obtain Eq.  (\ref{lambda-volume}). This is the reason why we have assumed shift symmetry in our 
setup.


\section{Cosmological Implications}\label{setup-sec}

In this section, we study cosmological implications of the two-field mimetic dark matter model
(\ref{action0}) at the background and perturbation levels. 

\subsection{Background equations}

We consider a spatially flat FRW background with spacetime metric  
\ba\label{FRWmetric}
ds^2 = - dt^2 + a(t)^2 d \bfx^2 \,,
\ea
where $a(t)$ is the scale factor and $t$ is the cosmic time. The mimetic constraint 
(\ref{mimetic-const.}) then implies 

\ba
\label{mimetic-const.BG}
\dot{\phi}^2+ {\cal S}^2 \dot{\psi}^2 = 1\,.
\ea

The Einstein's equations at the cosmological background give 
\ba
3 H^2 = - 2 \lambda \, ,
\ea
and
\ba\label{V0}
2 \dot H + 3 H^2 = 0\, ,
\ea
in which $H= \dot a(t)/a(t)$ is the Hubble expansion rate and also we have used the mimetic
constraint (\ref{mimetic-const.BG}). From the above equations, one can 
solve for $\lambda$, obtaining 
\ba
\label{lambda-eq0}
\lambda = \dot H \, .
\ea 
In addition, the modified Klein-Gordon equations 
(\ref{KG-eq}) give the following results
\ba\label{KG-eq-FRW}
\lambda a^3 {\dot\phi} = c_1\,,\hspace{1cm}
\lambda a^3 {\dot\psi} = c_2\,,
\ea
where $c_1$ and $c_2$ are some constants of integration. The constraint mimetic 
(\ref{mimetic-const.BG}) together with the modified Klein-Gordon equations (\ref{KG-eq-FRW}) imply 
\ba\label{lambda}
\lambda\propto{a^{-3}}\,.
\ea
The above result can be also obtained from Eq. (\ref{lambda-volume}) evaluated in 
cosmological background (\ref{FRWmetric}).
On the other hand, from the energy momentum tensor Eq. (\ref{Tmu-nu}), we can read the energy 
density and the pressure as $\rho=-T^0_0$ and  $P=\frac{1}{3}T^i_i$, which after substituting from 
(\ref{mimetic-const.BG}) and (\ref{lambda}), result in $\rho\propto{a^{-3}}$ and $P=0$.

Thus, although there are two scalar fields in our setup, but similar to the case of standard mimetic 
scenario \cite{Mimetic-2013}, it describes a fluid which behaves like the dark matter. In the next 
subsection, we make clear this apparent similarity with the case of single field by means of an 
appropriate decomposition of the scalar fields in the field space.

\subsection{Adiabatic and entropy decomposition}\label{AEdecomposition}

Comparing the constraint equation (\ref{mimetic-const.BG}) in our setup with its counterpart 
in single field scenario $\dot{\phi}^2 = 1$, we find that neither of the fields $\phi$ and $\psi$ 
individually play the role of the mimetic field in single field scenario. So, it is useful to consider 
a transformation in field space such that one of the new fields plays the role of mimetic field 
as in single field scenario.  In this respect, we can understand how the setup 
still describes dark matter-like fluid even in the presence of an extra scalar field. Following 
Ref. \cite{Wands:AEP}, we decompose $\phi$ and $\psi$ into the adiabatic $\sigma$ and 
entropy $s$ components through a rotation in field space as
\ba
\label{Field-decomposition}
\dot{\sigma}=(\cos{\theta})\, \dot{\phi}+(\sin{\theta})\, {\cal S}\dot{\psi}\,,
\ea
and
\ba
\label{Field-decomposition-entropy}
\dot{s}=-(\sin{\theta})\,\dot{\phi}+(\cos{\theta})\, {\cal S}\dot{\psi}\,,
\ea
where we have defined 
\ba\label{Field-decomposition2}
\cos{\theta} \equiv \frac{\dot{\phi}}{\sqrt{\dot{\phi}^2+{\cal S}^2\dot{\psi}^2}} \,,\hspace{2cm}
\sin{\theta} \equiv \frac{{\cal S}\dot{\psi}}{\sqrt{\dot{\phi}^2+{\cal S}^2\dot{\psi}^2}} \,.
\ea

Substituting Eq. (\ref{Field-decomposition2}) in Eq. (\ref{Field-decomposition}) and then again using
mimetic constraint (\ref{mimetic-const.BG}), we find
\ba\label{mimetic-C}
\dot{\sigma} = 1.
\ea
From the above relation it is clear that the field $\sigma$, which determines the path length 
along the classical trajectory, plays the role of mimetic field in single field scenario at the 
background level. We will see that it behaves the same as the mimetic field at the 
perturbations level as well.

Substituting Eq. (\ref{Field-decomposition2}) in Eq. (\ref{Field-decomposition-entropy}), yields
\be\label{s}
\dot{s} = 0 \, ,
\ee 
which shows that the entropy field is constant at the background level. This result is consistent with the expectation that at the background level there is no 
displacement in the direction perpendicular to classical trajectory \cite{Langlois:2006vv}. This constant  value $s$ is nothing but the parameter ${\cal S}$ which we have already defined in the previous section. To see this fact explicitly, note that if we set ${\cal S}$ to be zero, from (\ref{Field-decomposition2})  we have $\cos\theta=1$ and $\sin\theta=0$ which after substituting in (\ref{Field-decomposition})  gives $\dot\sigma=\dot{\phi}$. This result confirms that the parameter ${\cal S}$ is the value of entropy  field at the background. 
From now on, without loss of generality we absorb it into the field 
$\psi$ through the field redefinition $\psi\rightarrow\psi/{\cal S}$.

Using the mimetic constraint Eq. (\ref{mimetic-const.BG}) in Eq. (\ref{Field-decomposition2}), we find 
$\cos\theta=\dot{\phi}$ and $\sin\theta=s\dot{\psi}$. Taking the time derivative and then combining
the results, it is easy to show that
\ba\label{thetadot}
\dot{\theta} = s (\dot{\phi}\ddot{\psi}-\dot{\psi}\ddot{\phi}) = 0\,,
\ea
where in the last step we have used the fact that $\dot{\phi}\ddot{\psi}=\dot{\psi}\ddot{\phi}$ 
which can be deduced by taking the time derivative of the (\ref{KG-eq-FRW}).

Note that $\theta$ represents the rotation angle in field space and in general it can be time 
dependent. The above relation however shows that we deal with a constant rotation in field 
space. This is originated from the assumption that the model enjoys a shift symmetry in field 
space and there is no potential term. This changes when a potential term is added to the 
setup. 

At the level of perturbation, the fluctuation in scalar fields $\delta\phi$ and $\delta\psi$ are 
mapped to the adiabatic $\delta\sigma$ and entropic  $\delta{s}$ fluctuations given by
\ba\label{perturbations}
\delta\sigma&=&(\cos\theta)\delta\phi+(\sin\theta)\delta\psi\,,
\\ \nonumber
\delta{s}&=&-(\sin\theta)\delta\phi+(\cos\theta)\delta\psi\,.
\ea
In this view, $\delta\sigma$ represents the contribution of two fields perturbations 
$\delta\phi$ and $\delta\psi$  along the direction of background trajectory while 
$\delta{s}$ represents the fluctuations orthogonal to the classical trajectory. 


\subsection{Perturbations in comoving gauge}
\label{stability-sec}

In this section we present the cosmological perturbations analysis. To confirm that the results are not 
artifacts of specific gauge in which we are working, we perform the analysis in both comoving and 
spatially flat gauges. Here we present the analysis in comoving gauge while the analysis in flat gauge 
are relegated into the appendix \ref{app-flat}.  Moreover, we do not consider the coupling to the 
Standard Model fields in our model. Indeed, 
it is an open issue that to which metric the Standard Model fields is minimally coupled when one
performs a disformal/conformal transformation. For instance, if the matter minimally couples to the 
physical metric $g_{\mu\nu}$ in (\ref{disformal}), it should be non-minimally coupled to the auxiliary 
metric in the original frame and vice versa \cite{Zumalacarregui:2013pma}. In addition, as we will show 
later on, our model (\ref{action0}) obtained by performing a singular conformal transformation to the 
Einstein-Hilbert action, can provide a propagating scalar mode even in the absence 
of ordinary matter. Here,  we only perform the perturbation analysis of our model in the absence of 
ordinary matter in order to see whether or not the new propagating scalar mode in our model is free 
of disastrous pathologies.

In standard ADM decomposition, the metric perturbations are given by
\begin{equation}\label{ADM-metric}
ds^2=-N^2dt^2+h_{ij}(dx^i+N^i dt)(dx^j+N^j dt)\,,
\end{equation}
in which $N$ is the lapse function, $N^i$ are the components of shift vector, and $h_{ij}$ is the 
metric of the three-dimensional spatial part. In general, $h_{ij}$ contains two scalar degrees of 
freedom,  which after fixing one of them through choosing a gauge, it can be cast into the diagonal form 
\begin{equation}\label{hij}
h_{ij}=a^2 e^{2\psi_{(3)}}\delta_{ij}\,.
\end{equation}
On the other hand, the curvature perturbation is defined as $\calR \equiv \psi_{(3)} + 
\frac{H}{\dot{\sigma}} \delta \sigma$ which after imposing the mimetic constraint 
Eq. (\ref{mimetic-C}) becomes 
\ba\label{CPD}
\calR = \psi_{(3)} + H \delta\sigma\,.
\ea

Working in {\it comoving gauge} $\delta{\sigma}=0$, $\psi_{(3)}$ coincides with the curvature 
perturbation and we therefore set $\psi_{(3)} =\calR$ in the following analysis.

We are interested in scalar perturbations and we thus consider the first order scalar perturbations 
in metric such that\footnote{We use $B$ for both scalar perturbations in this Section and
coefficient of the disformal term in the previous Section. Since in the current Section the disformal 
coefficient is zero it does not cause confusion.}
\ba\label{ADM-scalar-pert}
N=1+N_1, \quad  N^i=\partial^i B,  \quad h_{ij}=a^2 
e^{2\calR}\delta_{ij}\,. 
\ea
Using the Guass-Codazzi relation $R= {^{(3)}R}+K_{ij}K^{ij}-K^2$ in which $^{(3)}R$ is the spatial 
curvature associated to the metric $h_{ij}$, $K_{ij}=\frac{1}{2N}(\dot{h}_{ij}-\nabla_i N_j
-\nabla_j N_i)$ is the extrinsic curvature, and $K=K^i_i$, and then substituting from 
Eq. (\ref{ADM-metric}) together with Eq. (\ref{ADM-scalar-pert}), it is straightforward to show
that the action (\ref{action0}) for the second order scalar perturbations takes the following form
\ba\label{action-2}
S^{(2)}_{\rm com}= \int d^{4}x\, a^3
\left(L_{EH}^{(2)} + {L}_M^{(2)}\right)\,,
\ea
in which $L_{EH}^{(2)}$ represents the contribution of the Einstein-Hilbert term given by
\ba\label{Lagrangian-EH}
L_{EH}^{(2)}
&=& -3\dot{\calR}^2-18H \calR\dot{\calR}-\dfrac{27}{2} H^2 \calR^2+
6H N_1 \dot{\calR}+ 9 H^2 N_1 \calR- 3 H^2 N_1^2
\\ \nonumber 
&-&\frac{1}{a^2}\Big(\left(\partial\calR\right)^2+2(N_1+\calR)
\partial^2\calR\Big)-2H (N_1-3\calR)\partial^2B+
6H\partial_i\calR \partial_iB+2\dot{\calR}\partial^2B
\\ \nonumber
&+& \partial_i\partial_jB\partial_i\partial_jB
-\left(\partial^2B\right)^2\,,
\ea
and ${ L}_M^{(2)}$ denotes the contribution from the mimetic matter fields which is given by
\ba\label{Lagrangian-M0}
L_M^{(2)}
&=& -\bar{\lambda}\left(\delta{\dot{s}}^2 + N_1^2-6 N_1 
\calR \right) + 2 \lambda^{(1)} N_1 \, ,
\ea
in which $\bar\lambda$ denotes the background value of the Lagrange multiplier $\lambda$ and 
$\lambda^{(1)}$ is its first order perturbation.

Going to Fourier space and doing some integration by parts, we obtain the following 
Lagrangian density for the second order action\footnote{Note that we do not write the 
dependence of perturbations on Fourier wave number $k$ and $\calR(k)$ is simply denoted 
by $\calR$ and so on.} Eq. (\ref{action-2})\ba\label{Lagrangian}
{\cal L}^{(2)}_{\rm com}&=&
\frac{3}{2} a^3 H^2 \delta{\dot{s}}^2 - \frac{3}{2} a H^2 k^2 
\delta{s}^2-3 a^3 \dot{\calR}^2+2 a^3 (3HN_1-k^2B) 
\dot{\calR} + ak^2{\calR^2}
\\ \nonumber
&+&  2 a k^2 N_1 R -\frac{3}{2} a^3 H^2 N_1^2 
+ 2 a^3 H k^2 B N_1+2 a^3 \lambda^{(1)} N_1
\ea
where we have substituted $\dot{H}=-\frac{3}{2} H^2$ and $\lambda=\dot{H}$ from 
Eqs. (\ref{V0}) and (\ref{lambda-eq0}) respectively.

The equation of motion for $\lambda^{(1)}$ and $B$ from the Lagrangian 
Eq. (\ref{Lagrangian}) lead to the following two constraints
\ba\label{EoM-lambda}
N_1 = 0\,,
\ea
and 
\ba\label{EoM-alpha}
\dot{\calR}=0\,.
\ea
The above relation shows that the curvature perturbation $\calR$ does not propagate in our 
setup. This result is similar to the case of single field mimetic matter scenario \cite{Mimetic-2013}.

Plugging the above results into the equation of motion for $N_1$, we obtain the following 
solution for $B$ 
\ba\label{EoM-psi}
B = -\frac{\calR}{a^2 H} - \frac{\lambda^{(1)}}{k^2 H}\,.
\ea
Substituting the above results in (\ref{Lagrangian}), the reduced Lagrangian for the second
order perturbation in comoving gauge is obtained to be
\ba\label{LagrangianR}
{\cal L}^{(2)}_{\rm com}&=&
\frac{3}{2} a^3 H^2 \delta{\dot{s}}^2 - \frac{3}{2} a H^2 k^2 \delta{s}^2
+a k^2 \calR^2\,.
\ea
In order to study the stability of the setup, we should obtain the Hamiltonian. The 
associated canonical momenta are given by $\Pi_{\calR}=0$ and $\Pi_{\delta{s}}=3 a^3 H^2 
\delta{\dot{s}}$. So, we have to implement the primary constraint $\Pi_\calR=0$ which leads 
to the secondary constraint $\calR=0$ through the consistency condition ${\dot{\Pi}}_\calR
=0$. More precisely, both of the constraints are second class and therefore the total 
number of physical degrees of freedom is one which is $\delta{s}$ (the phase space is 
two-dimensional). 

After imposing the constraints, the reduced Hamiltonian is given by 
\ba\label{HamiltonianR}
{\cal H}^{(2)}_{\rm com}&=&
\frac{\Pi_{\delta{s}}^2}{6 a^3 H^2}+\frac{3}{2} a H^2 k^2 \delta{s}^2\,.
\ea
From the above Hamiltonian function, it is clear that there is only one propagating mode, 
$\delta{s}$, which is healthy, propagating   with the speed of unity. 

Note that, as in  original mimetic model, the adiabatic mimetic mode $\sigma$ is 
non-propagating. This is in line with the fact that the mimetic background describes a 
fluid with no pressure so one expects the sound speed for the adiabatic mode to be zero. 
As a result there is no notion of quantum wave describing the mimetic field perturbations. 
It is expected that  the perturbations in the adiabatic mimetic field with no pressure to 
generate caustic instabilities in dark matter perturbations  so the two-field mimetic 
setup with  zero sound speed  may not be appealing.  However look at 
\cite{DeFelice:2015moy, Gumrukcuoglu:2016jbh} and 
\cite{Babichev:2016jzg, Babichev:2017lrx} where it was  argued that this may not be a 
serious problem. On the other hand, in the two-dimensional field space, the perturbations 
perpendicular to background trajectory is excited and can be used in cosmological 
applications of mimetic scenario.


\section{Discussions}
\label{summary-sec}

The mimetic gravity scenario can be uniquely obtained from the singular limit of disformal
transformation. Therefore, in order to find the two-field extension of the standard mimetic gravity, 
we have extended the disformal transformation to the case of two scalar fields. The most general 
form of the two-field disformal transformation (\ref{dis}) would contain a cross term between two 
scalar fields labeled by the coefficient $D$. Performing two successive disformal transformations
cannot generate this cross term. However, we have shown that this cross term can be removed 
through a one-to-one linear transformation in cotangent space of the field space. This  shows 
that the most general two-field disformal transformation is equivalent to two successive single 
field disformal transformations. We then studied the transformation between the ``physical'' and 
the original ``auxiliary" metrics through the Jacobian of the transformation. Solving the 
corresponding eigentensor equation, we have found the associated eigentensors and eigenvalues.
We then looked at the singular limit of the conformal two-field transformation as the two-field 
generalization of the mimetic scenario. 

At the cosmological background, the setup describes a dark matter-like fluid
much similar to the standard single field mimetic scenario. However, as expected, they differ at the 
perturbation level. Decomposing the modes into the adiabatic and entropy components, we have 
found that, similar to the standard single field mimetic model, the adiabatic mode does not 
propagate in this model. But, the entropy mode, originating from the extra scalar field in our setup, 
propagates with speed of unity and is free of any disastrous pathologies. In order to make sure that 
these results are not artifacts of any particular gauge which one uses, we have  performed the 
perturbations analysis in both comoving and spatially flat gauges.

There is a number of directions in which the current analysis can be extended. The first direction is to 
consider $N>2$ multiple fields mimetic setup. For this purpose, one has to extend the disformal 
transformation to $N$ fields and then look for its singular limit. The eigenvalue and the eigentensor 
analysis for the general disformal transformation are expected to be very complicated. However, as in 
the current work, much insights can be obtained if one looks at the conformal limit. The second 
direction is to break the assumption of shift symmetry and allow a potential term $V(\phi, \psi)$ in the 
constrained Lagrangian Eq.  (\ref{action0}). The experience with the single field mimetic setup indicates 
that the adiabatic mode is no longer frozen. Furthermore, it is expected to suffer from pathologies 
such as the ghost and gradient instabilities. To remedy these pathologies, as in standard mimetic 
scenario, one may need to couple the  higher derivatives of mimetic fields to curvature terms. It is an 
interesting exercise to see if one can get healthy propagating adiabatic modes by coupling the higher 
derivatives of the mimetic fields to curvature terms when the shift symmetry is broken.

\vspace{1cm}


{\bf Acknowledgments:}  We would like to thank Nathalie Deruelle, Shinji Mukohyama, Borna Salehian  
and David Wands for useful discussions and correspondences. H. F., M. A. G. and A. K.  thank the 
Yukawa Institute for Theoretical Physics at Kyoto University for hospitality.  Discussions during the 
YITP symposium YKIS2018a ``General Relativity -- The Next Generation --" were useful to complete 
this work. We also thank the anonymous referee for insightful comments which improved the presentations of the draft.

\vspace{0.7cm}

\appendix
\renewcommand{\theequation}{A-\arabic{equation}}

\section{Diagonalizing two-field disformal transformation}
\label{app-diag}

In this Appendix, our aim is to show that it is always possible to remove the off-diagonal term, 
controlled by the coefficient  $D$, in the two-field disformal transformation (\ref{dis}) by means 
of an appropriate transformation. 

We therefore consider the following linear map in cotangent space of the field space
\begin{align}\label{transformation-cotangent}
& d\phi = \alpha_1\, d\chi + \alpha_2\, d\eta \,, \nonumber \\
& d\psi = \alpha_3\, d\chi + \alpha_4\, d\eta \,,
\end{align}
in which $\alpha_i$ are functions of $\phi$, $\psi$, $X$, $Y$, and $Z$ which are defined as
\[
\begin{array}{cc} \alpha_1 \equiv  \frac{\partial\phi}{\partial\chi} \,, \quad 
& \alpha_2 \equiv  \frac{\partial\phi}{\partial\eta}  \,, \\ \alpha_3 \equiv \frac{\partial\psi}{\partial\chi}  \,, \quad 
& \alpha_4 \equiv \frac{\partial\psi}{\partial\eta}  \,. \end{array}
\]

In matrix notation, transformation (\ref{transformation-cotangent}) can be rewritten as
\begin{equation}
\left(\begin{array}{c} d\phi \\ d\psi \end{array}\right) 
=
\left(\begin{array}{cc} \alpha_1 & \alpha_2 \\ \alpha_3 & \alpha_4 \end{array}\right)
\left(\begin{array}{c} d\chi \\ d\eta \end{array}\right) \,,
\end{equation}
and to have invertible transformation, we demand that the determinant of the transformation
matrix to be  nonzero
\be
\det\alpha= \alpha_1 \alpha_4 - \alpha_2 \alpha_3 \neq 0  \,.
\ee

In component form, transformation (\ref{transformation-cotangent}) also implies
\begin{align}\label{transformation-component}
& \phi_{,\mu} = \alpha_1\, \chi_{,\mu} + \alpha_2\, \eta_{,\mu} \,, 
\nonumber \\
& \psi_{,\mu} = \alpha_3\, \chi_{,\mu} + \alpha_4\, \eta_{,\mu} \,.
\end{align}

Substituting (\ref{transformation-component}) into the two-field disformal transformation
(\ref{dis}), we find 
\be
\label{dis-new}
g_{\mu\nu} = A {\tilde g}_{\mu\nu}
+ \tilde{B} \chi_{,\mu} \chi_{,\nu} + \tilde{C} \eta_{,\mu} \eta_{,\nu} 
+ \tilde{D} (\chi_{,\mu} \eta_{,\nu} + \eta_{,\mu} \chi_{,\nu})  \,,
\ee
in which we have defined
\begin{eqnarray}\label{new-coefficient}
&&{\tilde B} \equiv \alpha_1^2 B + \alpha_3^2 C + 2 \alpha_1 \alpha_3 D  \,, \\ \nonumber
&&{\tilde C} \equiv \alpha_2^2 B + \alpha_4^2 C + 2 \alpha_2 \alpha_4 D  \,, \\ \nonumber
&&{\tilde D} \equiv  \alpha_1 \alpha_2 B + \alpha_3 \alpha_4 C
+ (\alpha_1 \alpha_4 + \alpha_2 \alpha_3) D  \,.
\end{eqnarray}

The coefficients $A, {\tilde B}, {\tilde C}, {\tilde D}$ are now functions of $\chi$, $\eta$, 
${\tilde X}$, ${\tilde Y}$, and ${\tilde Z}$ which are defined as 
\be
\left\{
\begin{array}{c}
	{\tilde X} = {\tilde g}^{\mu\nu}\partial_\mu\chi\partial_\nu\chi  \,, \\
	{\tilde Y} = {\tilde g}^{\mu\nu}\partial_\mu\eta\partial_\nu\eta  \,, \\
	{\tilde Z} = {\tilde g}^{\mu\nu}\partial_\mu\chi\partial_\nu\eta  \, ,
\end{array}
\right.
\ee
which are linearly related to their old counterparts as follows 
\begin{eqnarray}
&& {\tilde X} = (\det\alpha)^{-2} \left( \alpha_4^2 X + \alpha_2^2 Y 
- 2 \alpha_2 \alpha_4 Z \right)  \,, \\ \nonumber
&& {\tilde Y} = (\det\alpha)^{-2} \left( \alpha_3^2 X + \alpha_1^2 Y 
- 2 \alpha_1 \alpha_3 Z \right)  \,, \\ \nonumber
&& {\tilde Z} = (\det\alpha)^{-2} \big( \alpha_3 \alpha_4 X + \alpha_1 \alpha_2 Y 
- (\alpha_2 \alpha_3 + \alpha_1 \alpha_4) Z \big) \,.
\end{eqnarray}

In order to remove the off-diagonal term, we demand that $\tilde{D}=0$ in (\ref{new-coefficient}),  
which after solving for $\alpha_4$, gives
\be\label{alpha4}
\alpha_4 = - \left(\frac{\alpha_1 B + \alpha_3 D}{\alpha_1D+ \alpha_3 C}\right) \alpha_2 \,.
\ee

Substituting the above solution in (\ref{dis-new}) we can remove the off-diagonal term and
therefore we are left with the diagonal two-field disformal transformation
\be
\label{dis-diagonal}
g_{\mu\nu} = A {\tilde g}_{\mu\nu}
+ \tilde{B} \chi_{,\mu} \chi_{,\nu} + \tilde{C} \eta_{,\mu} \eta_{,\nu}  \,.
\ee
Let us consider the simple case of $\det\alpha=1$ and $\alpha_3=0$ and $\alpha_1=1$ which 
implies $\alpha_4=1$ and $\alpha_2 = - D/B $. Substituting these particular choices in 
(\ref{transformation-component}), we can easily find
\begin{align}\label{transformation-component-particular}
& \phi_{,\mu} = \chi_{,\mu} -  (D/B) \, \eta_{,\mu}  \,, \nonumber \\
& \psi_{,\mu} = \eta_{,\mu}  \,.
\end{align}

It is easy to directly check that the above simple linear transformation diagonalizes the two-field
disformal transformation (\ref{dis}) as
\be
\label{dis-diagonal-particular}
g_{\mu\nu} = A \, {\tilde g}_{\mu\nu}
+ B \, \chi_{,\mu} \chi_{,\nu} + \left( C - \frac{D^2}{B} \right) \eta_{,\mu} \eta_{,\nu} \,.
\ee

Therefore, we can always remove the off-diagonal term in (\ref{dis}) such that it takes the 
diagonal form (\ref{dis-diagonal}). But, from (\ref{coefficient-solution}) it is clear that $\bar{D}$
is nonzero even if we set $D=0$. This means that the inverse metric $\tilde g^{\mu \nu}$ in 
(\ref{metric-inverse}) would have an off-diagonal term even if we start with a diagonal form 
in $\tilde g_{\mu \nu}$. 

Moreover, if we apply the transformation (\ref{transformation-component}) into 
Eq. (\ref{singular-solution}) through identifying $B$, $C$ and $D$ with $\alpha$, $\beta$, and 
$\gamma$ respectively, it is straightforward to show that the term proportional to $Z$ can be 
removed. For simplicity, we work with the particular case 
(\ref{transformation-component-particular}), which after the above mentioned identification of
the coefficients, transforms Eq. (\ref{singular-solution}) into the following canonical form
\be
A = \alpha \tilde{X} + \left( \beta - \frac{\gamma^2}{\alpha} \right) \tilde{Y} \,.
\ee
So without loss of generality, we can set $\gamma =0$ in Eq. (\ref{singular-solution}). 


\section{Eigenvalue problem for the most general two-field disformal transformation}
\label{app-general-dis}

In this Appendix, we solve the eigenvalue problem Eq. (\ref{eigen}) in its most general form. In 
the case of two-field disformal transformation Eq. (\ref{dis}), Eq. (\ref{eigen}) takes the 
following form

\be
\label{eigenG}
(A-\lambda)\xi_{\mu\nu}-{\mathcal M}_{\mu\nu}^{\alpha\beta}\xi_{\alpha\beta} = 0 \,,
\ee
where we have defined 

\begin{align}
\label{M}
{\mathcal M}_{\mu\nu}^{\alpha\beta} &= 
\Big(A_{,X} {\tilde g}_{\mu\nu}+B_{,X}\phi_{,\mu}\phi_{,\nu}
+C_{,X}\psi_{,\mu}\psi_{,\nu}+D_{,X}(\phi_{,\mu}\psi_{,\nu}
+\psi_{,\mu}\phi_{,\nu})\Big)
\phi^{,\alpha}\phi^{,\beta} \nonumber\\
&+\Big(A_{,Y}{\tilde g}_{\mu\nu}+B_{,Y}\phi_{,\mu}\phi_{,\nu}
+C_{,Y}\psi_{,\mu}\psi_{,\nu}
+D_{,Y}(\phi_{,\mu}\psi_{,\nu}+\psi_{,\mu}\phi_{,\nu})\Big)
\psi^{,\alpha}\psi^{,\beta} \\
&+\Big(A_{,Z}{\tilde g}_{\mu\nu}+B_{,Z}\phi_{,\mu}\phi_{,\nu}
+C_{,Z}\psi_{,\mu}\psi_{,\mu}
+D_{,Z}(\phi_{,\mu}\psi_{,\nu}+\psi_{,\mu}\phi_{,\nu})\Big)
\phi^{,\alpha}\psi^{,\beta} \,. \nonumber
\end{align}

The above equation is also the equation for eigenvalues and eigentensors of 
$\mathcal M_{\mu\nu}^{\alpha\beta}$. In order to solve the above eigenvalue equation, we note 
that we deal with the space of all symmetric $4\times4$ matrices which can be spanned by 
means of ten tensor basis $e^{(i)}_{\mu\nu}$ with $i=1,..,10$. Working with orthonormal 
orthogonal basis, we have
\be\label{orthogonality}
e^{(i)}_{\mu\nu}\, e_{(j)\mu\nu} = \delta^i_j \,.
\ee
Therefore, we can expand ${\tilde g}_{\mu\nu}, \phi_{,\mu} \phi_{,\nu}, \psi_{,\mu}\psi_{,\nu}$, 
and $\phi_{,\mu}\psi_{,\nu}$ in terms of these basis as \footnote{Note that we can expand all 
$4\times4$ matrices in terms of tetrads. But, here we just deal with the symmetric subset of 
all $4\times4$ matrices and we prefer to work with ten tensor basis (\ref{orthogonality}) which 
are sufficient and also more appropriate for our purpose in this paper.}
\be
\label{zeta}
\zeta_{\mu\nu}^I=\sum_{i=1}^{10} c^I_i e^{(i)}_{\mu\nu} \,,
\ee
where $I = 1, 2, 3, 4$ corresponds to ${\tilde g}_{\mu\nu}, \phi_{,\mu} \phi_{,\nu},
\psi_{,\mu}\psi_{,\nu}$, and $\phi_{,\mu}\psi_{,\nu}$ respectively and $c^I_i$ are clearly the 
associated components. If we fix the explicit form of the basis then we can find the explicit 
form of the components $c^I_i$. However, as we shall see, we do not need to fix the explicit 
form of basis. 

In the same way, the eigentensors can be expanded in terms of basis (\ref{orthogonality}) as 
follows
\be
\label{xi}
\xi_{\mu\nu}=\sum_{i=1}^{10} a_i e^{(i)}_{\mu\nu} \,,
\ee
where $a_i$ are the associated components. 

Substituting Eq. (\ref{zeta}) into Eq. (\ref{M}), we can express ${\cal M}_{\mu\nu}^{\alpha\beta}$ 
in terms of the basis Eq. (\ref{orthogonality}) as
\be\label{M-component}
\mathcal M_{\mu\nu}^{\alpha\beta}=\sum_{i,j=1}^{10}\sum_{I,J=1}^{4} M_{IJ} c^I_i c^{Jj} 
e^{(i)}_{\mu\nu}e^{\alpha\beta}_{(j)}\,,
\ee
in which we have defined $4\times4$ matrix $M_{IJ}$ as follows

\[
{\boldmath M}_{IJ} =
\left[ {\begin{array}{cccc}
	0 & A_{,X} & A_{,Y} & A_{,Z} \\
	0 & B_{,X} & B_{,Y} & B_{,Z} \\
	0 & C_{,X} & C_{,Y} & C_{,Z} \\
	0 & 2 D_{,X} & 2 D_{,Y} & 2 D_{,Z} \\
	\end{array} } \right] \,.
\]

Substituting Eq. (\ref{M-component}) together with Eq. (\ref{xi}) into the eigenvalue equation 
(\ref{eigenG}) and using orthogonality condition Eq. (\ref{orthogonality}) give

\be\label{eigen-basis}
\sum_{i=1}^{10} e^{(i)}_{\mu\nu} \Big[(A-\lambda)a_i
- \sum_{j=1}^{10}\sum_{I,J=1}^{4}M_{IJ} c^{Jj} a_j c^I_i \Big] = 0 \,.
\ee

From the above eigenvalue equation, it is clear that the conformal type eigenvalue is again a solution 
with
\be\label{eigenvalue-c}
\lambda^{C} = A \,, \hspace{1cm} \mbox{with}
\hspace{1cm} \sum_{i,j=1}^{10}\sum_{I,J=1}^{4}M_{IJ} c^{Jj} a_j c^I_i e^{(i)}_{\mu\nu} = 0  \,.
\ee
Note that this imposes one constraint on the eigentensors and therefore the conformal type eigenvalue 
is degenerate with multiplicity of $9$.

For the remaining eigenvalue, we note that Eq. (\ref{eigen-basis}) can be satisfied for the eigentensor 
$\xi_{\mu\nu}=\,\sum_{I=1}^4 \tilde{a}_I \zeta_{\mu\nu}^I=\sum_{I=1}^4 \sum_{i=1}^{10} \tilde{a}_I 
c^I_i e^{(i)}_{\mu\nu}$ in which ${\tilde a}_I$ are components of $a_i$ in direction of $c^I_i$ (we have 
used Eq. (\ref{zeta}) as well). Substituting this ansatz into Eq. (\ref{eigen-basis}), we find that 
${\tilde a}_{I}{\propto} \sum_{j=1}^{10}\sum_{I,J,K=1}^{4}M_{IJ} c^{Jj} {\tilde a}_K c^K_j $. 
Therefore, the kinetic type eigenvalue and the corresponding eigentensor in Eq. (\ref{eigen-basis}) 
will be
\be\label{eigenvalue-k}
\lambda^{\cal K} = A - a \,, \hspace{1cm} \mbox{with}
\hspace{1cm} \xi^{\cal K}_{\mu\nu} 
= a \sum_{j=1}^{10}\sum_{I,J,K=1}^{4}M_{IJ} c^{Jj} {\tilde a}_K c^K_j c^I_i e^{(i)}_{\mu\nu}  \,,
\ee
where $a$ is an unknown function of $\phi, \psi, X, Y, Z$. 

Our aim is now to find the explicit form of the kinetic eigenvalue or equivalently to find the explicit form 
of $a$. In order to do this, we note  that the kinetic eigentensor is aligned in direction of $c^I_i$ as 
$\xi_{\mu\nu}^{\cal K}=\,\sum_{I=1}^4 \tilde{a}_I \zeta_{\mu\nu}^I=\sum_{I=1}^4 \sum_{i=1}^{10} 
\tilde{a}_I c^I_i e^{(i)}_{\mu\nu}$. Therefore, it is clear that $\xi^{\cal K}_{\mu\nu} \propto {a} 
{\tilde g}_{\mu\nu} + b \phi_{,\mu} \phi_{,\nu} + c \psi_{,\mu} \psi_{,\nu} 
+ d (\phi_{,\mu} \psi_{,\nu} + \psi_{,\mu} \phi_{,\nu} )$ in which
\begin{align}\label{abcd}
a =\, & A_{,X} \langle\xi^{\cal K}\rangle_{X} + A_{,Y} \langle\xi^{\cal K}\rangle_{Y} + A_{,Z} 
\langle\xi^{\cal K}\rangle_{Z}\,, \nonumber \\  
b =\, & B_{,X} \langle\xi^{\cal K}\rangle_{X} + B_{,Y} \langle\xi^{\cal K}\rangle_{Y} + B_{,Z} 
\langle\xi^{\cal K}\rangle_{Z}\,, \nonumber \\  
c =\, & C_{,X} \langle\xi^{\cal K}\rangle_{X} + C_{,Y} \langle\xi^{\cal K}\rangle_{Y} + C_{,Z} 
\langle\xi^{\cal K}\rangle_{Z}\,, \nonumber \\  
d =\, & D_{,X} \langle\xi^{\cal K}\rangle_{X} + D_{,Y} \langle\xi^{\cal K}\rangle_{Y} + D_{,Z} 
\langle\xi^{\cal K}\rangle_{Z} \,,
\end{align}
are defined from Eq. (\ref{eigen}). We show that $a$ in the above relation coincides with what is 
already defined in Eq. (\ref{eigenvalue-k}). The normalization factor is however important since the 
coefficient $a$ is defined in terms of $\langle\xi^{\cal K}\rangle_{X}$, 
$\langle\xi^{\cal K}\rangle_{Y}$, and $\langle\xi^{\cal K}\rangle_{Z}$ in Eq. (\ref{abcd}). We 
therefore consider the following combination

\be\label{xi-total}
\xi^{\cal K}_{\mu\nu} = {\tilde g}_{\mu\nu} 
+ \frac{b}{a} \phi_{,\mu} \phi_{,\nu} + \frac{c}{a} \psi_{,\mu} \psi_{,\nu} 
+ \frac{d}{a} ( \phi_{,\mu} \psi_{,\nu} + \psi_{,\mu} \phi_{,\nu} ) \,.
\ee
In order to find the explicit form of $a, b, c, d$, we need to determine the various components in 
Eq. (\ref{xis-def}) in the case of (\ref{xi-total}). We therefore contract (\ref{xi-total}) with 
$\phi^{,\mu} \phi^{,\nu}$, $\psi^{,\mu} \psi^{,\nu}$ and $\phi^{,\mu} \psi^{,\nu}$ which give
\begin{align}\label{xis}
a \langle\xi^{ K}\rangle_{X} =\, & a X + b X^2 + c Z^2 + 2 d X Z \nonumber \\  
a \langle\xi^{ K}\rangle_{Y} =\, & b Y + b Z^2 + c Y^2 + 2 d Y Z \nonumber \\  
a \langle\xi^{ K}\rangle_{Z} =\, & a Z + b X Z + c Y Z + d ( X Y + Z^2 ) \,.
\end{align}
These are algebraic second order equations which can be solved to obtain the explicit solutions
for $\langle\xi^{ K}\rangle_{X}$, $\langle\xi^{ K}\rangle_{Y}$, and 
$\langle\xi^{ K}\rangle_{Z}$. After finding them, the explicit form of $a, b, c, d$ will be 
determined. We do not write the explicit forms of the kinetic type eigenvalues and their associated 
eigentensors since they have messy expressions.

For the conformal case with $b=c=d=0$, Eq. (\ref{xi-total}) becomes 
$\xi^{ K}_{\mu\nu}={\tilde g}_{\mu\nu}$. From Eq.  (\ref{xis}) we obtain $\langle\xi^{ K}\rangle_{X} 
= X$, $\langle\xi^{ K}\rangle_{Y} = Y$, and $\langle\xi^{ K}\rangle_{Z} = Z$ which after 
substituting in (\ref{abcd}) yields $a = X A_{,X} + Y A_{,Y} + Z A_{,Z}$ and the 
corresponding eigenvalue Eq. (\ref{eigenvalue-k}) correctly coincides with Eq. (\ref{eigenvalue-Kphipsi}).

For the single field case with $\psi\equiv0$, from Eq. (\ref{xis}) we have $\langle\xi^{ K}\rangle_{X} 
= X + \big(\frac{b}{a}\big) X^2$ and the coefficients $a$ and $b$ are given by $a = A_{,X} 
\langle\xi^{ K}\rangle_{X}$ and $b = B_{,X} \langle\xi^{ K}\rangle_{X}$ through their 
definitions Eq. (\ref{abcd}). The eigentensor (\ref{xi-total}) then turns out to be 
$\xi^{ K}_{\mu\nu} = {\tilde g}_{\mu\nu} + \big(\frac{B_{,X}}{A_{,X}}\big) \phi_{,\mu} \phi_{,\nu}$ 
and the corresponding eigenvalue can be read from (\ref{eigenvalue-k}) as $\lambda^{ K}= 
A - X A_{,X} - X^2 B_{,X}$ in agreement with the results of \cite{Zumalacarregui:2013pma}.

\section{Perturbations in spatially flat gauge}
\label{app-flat}

In this appendix we present the cosmological perturbation analysis in spatially flat gauge. 
We show that the results are consistent  with those obtained in the comoving gauge. 

In spatially flat gauge  $\psi_{(3)}=0$ and therefore $h_{ij}=a^2\delta_{ij}$. The perturbed 
metric then takes the following simple form
\begin{equation}\label{metric-SF}
N=1+N_1, \quad  N^i=\partial^i B,   
\end{equation}
in which, as before,  $N_1$ and $B$ characterize the scalar perturbations in metric. For the matter 
part, there are two other scalar perturbations $\delta{s}$ and $\delta{\sigma}$. Substituting 
(\ref{metric-SF}) in (\ref{action0}), it is straightforward to show that the quadratic action is
\ba\label{action-2-SF}
S^{(2)}_{\rm flat}= \int d^{4}x\, a^3
\left(L_{EH}^{(2)} + {L}_M^{(2)}\right)\,,
\ea
in which $L_{EH}^{(2)}$ represents the contribution of the Einstein-Hilbert term in spatially flat 
gauge 
\ba\label{Lagrangian-EH-SF}
L_{EH}^{(2)}
&=& - 3 H^2 N_1^2 - 2H N_1 \partial^2B
+ \partial_i\partial_jB\partial_i\partial_jB
-\left(\partial^2B\right)^2\,,
\ea
and $L_M^{(2)}$ denotes the contribution of the matter part
\ba\label{Lagrangian-M0-SF}
L_M^{(2)}
&=& -\bar{\lambda}\left(\delta{\dot{s}}^2+\delta{\dot{\sigma}}^2 
+ N_1^2-2 N_1 \delta{\dot{\sigma}} - 2 \partial_i\delta{\sigma}
\partial_i B - a^{-2}(\partial \delta{s})^2 - 
a^{-2}(\partial \delta{\sigma})^2 \right) 
\nonumber \\
&+& 2 \lambda^{(1)} (N_1-\delta{\dot{\sigma}}) \, .
\ea

Going to Fourier space, we obtain the following reduced Lagrangian for the 
second order action (\ref{action-2-SF})
\ba\label{Lagrangian-SF}
{\cal L}^{(2)}_{\rm flat}&=&
\frac{3}{2} a^3 H^2 \delta{\dot{s}}^2 - \frac{3}{2} a H^2 k^2 
\delta{s}^2 + \frac{3}{2} a^3 H^2 \delta{\dot{\sigma}}^2 - 
\frac{3}{2} a H^2 k^2 \delta{\sigma}^2
\\ \nonumber
&-& \frac{3}{2} a^3 H^2 N_1^2 - 3 a^3 H^2 N_1 
\delta{\dot{\sigma}} + 2 a^3 H k^2 B N_1 - 3a^3H^2k^2
B\delta{\sigma} + 2 a^3 \lambda^{(1)} (N_1-\delta{\dot{\sigma}}) \,,
\ea
where again we have substituted $\dot{H}=-\frac{3}{2} H^2$ and $\lambda=\dot{H}$ 
from Eqs. (\ref{V0}) and (\ref{lambda-eq0}).

The equation of motion for $\lambda^{(1)}$ then gives
\ba\label{EoM-lambda-SF}
N_1 = \delta{\dot \sigma}\,.
\ea
Substituting this into (\ref{Lagrangian-SF}), the resultant Lagrangian gives the following 
equation of motion for the variation of the $B$ field 
\ba\label{EoM-psi-SF}
\delta{\dot \sigma}=\frac{3}{2} H \delta\sigma \, .
\ea
From the definition  (\ref{CPD}), the curvature perturbation in spatially flat gauge is given by $\calR=
H \delta\sigma$ which from (\ref{EoM-psi-SF}) we conclude $\dot{\calR}=0$ as before.

Substituting the above results into the Lagrangian (\ref{Lagrangian-SF}) gives
\ba\label{LagrangianR-SF}
{\cal L}^{(2)}_{\rm flat}&=&
\frac{3}{2} a^3 H^2 \delta{\dot{s}}^2 - \frac{3}{2} a H^2 k^2 \delta{s}^2
- \frac{3}{4} a H^2 \left(2k^2+9a^2H^2\right) \delta{\sigma}^2\,.
\ea
Going to the Hamiltonian formalism, the associated canonical momenta are given by 
$\Pi_{\delta\sigma}=0$ and $\Pi_{\delta{s}}=3 a^3 H^2 \delta{\dot{s}}$. Therefore, 
$\Pi_{\delta{\sigma}}=0$ is a primary constraint which generates the secondary constraint 
$\delta{\sigma}=0$ through the consistency condition ${\dot \Pi}_{\delta{\sigma}}=0$. It is
not difficult to show that after imposing the constraints, the resultant reduced Hamiltonian 
 coincides exactly with Eq. (\ref{HamiltonianR}) that we have obtained in comoving gauge in 
 subsection \ref{stability-sec}. The stability analysis is therefore the same as in subsection 
\ref{stability-sec} and this can be seen as a consistency check of our calculations.

{}

\end{document}